\documentclass[twocolumn,aps,pre,showpacs,superscriptaddress]{revtex4}
\usepackage[utf8]{inputenc}
\usepackage[T1]{fontenc}
\usepackage{graphicx}
\usepackage{amsmath,amssymb,amsfonts,array,subfigure}
\usepackage{float}
\usepackage{xcolor}
\begin{document}
\title{Dust-acoustic wave electrostatic and self-gravitational potentials in an opposite polarity dusty plasma system}
\author{Abdul Mannan}
\email{abdulmannan@juniv.edu}
\affiliation{Department of Physics, Jahangirnagar University, Savar, Dhaka-1342, Bangladesh}
\affiliation{Institut f\"{u}r Mathematik, Martin Luther Universit\"{a}t Halle-Wittenberg, D-06099 Halle (Saale), Germany}
\author{Sergio De Nicola}
\affiliation{SPIN-CNR, Complesso Universitario di M.S. Angelo, Napoli, Italy}
\affiliation{INFN Sezione di Napoli, Complesso Universitario di M.S. Angelo, Napoli, Italy}
\affiliation{Dipartimento di Fisica ``E. Pancini", Universit\`{a} di Napoli Federico II, Complesso Universitario di M.S. Angelo, via Cintia, I-80126
Napoli, Italy}
\author{Renato Fedele}
\affiliation{Dipartimento di Fisica ``E. Pancini", Universit\`{a} di Napoli Federico II, Complesso Universitario di M.S. Angelo, via Cintia, I-80126
Napoli, Italy}
\affiliation{INFN Sezione di Napoli, Complesso Universitario di M.S. Angelo, Napoli, Italy}
\author{A A Mamun}
\altaffiliation[Also at~]{Wazed Miah Science Research Centre, Jahangirnagar University, Savar, Dhaka-1342, Bangladesh.}
\affiliation{Department of Physics, Jahangirnagar University, Savar, Dhaka-1342, Bangladesh}
\date{\today}

\begin{abstract}
An opposite polarity dusty plasma system (containing a few micron size massive opposite polarity dust species and singly charged ion species following Boltzmann law) is considered. The nature of dust-acoustic (DA) wave electrostatic and self-gravitational potentials are correctly found by the numerical analysis of two coupled second-order nonlinear differential equations for electrostatic and self-gravitational potentials associated with the DA waves in such an opposite polarity dusty plasma medium. These coupled nonlinear differential equations are derived from the continuity and momentum equations for positive and negative dust species, and the Boltzmann law for ion species. The basic features of the DA wave self-gravitational potential are compared with that of the DA wave electrostatic potential. The relevance of our results to space and laboratory opposite polarity dusty plasma systems is mentioned.
\end{abstract}
\maketitle
\section{Introduction}
After the confirmation of the existence of opposite polarity dust species (OPDS) in space \cite{Havnes96,Tsintikidis96,Horanyi96,Gelinas98} and  laboratory devices \cite{Ali98,DAngelo01,DAngelo02}, the opposite polarity dusty plasma (OPDP) systems have received a great deal of interest in understanding the physics of electrostatic and self-gravitational perturbation in space environments (viz.  Earth's mesosphere \cite{Havnes96,Gelinas98}, cometary tails \cite{Horanyi96}, Saturn rings \cite{Tsintikidis96,Horanyi96}, etc.) and different laboratory devices  \cite{Ali98,DAngelo01,DAngelo02}. The most important property of this OPDP, which makes the latter unique and different from other types of pair plasmas (viz. electron-ion and electron-positron plasmas), is that the ratio
($\varrho$) of the size of positive dust species to that of negative dust species can be smaller or larger than unity, even it can
be equal to unity \cite{Chow93,Medis94,Medis04,Shukla06}. The size of the dust species is very important because of its important role in determining the mass and charge of dust species in situations of both positive and negative dust species \cite{Chow93,Medis94,Medis04,Shukla06}.

The OPDP systems have been considered by a large number of authors  \cite{Mamun02,Shukla05,Sayed07,Rahman08,El-Labany08,Mamun08a,Mamun08b,Verheest09,Mamun11,Mannan11,El-Taibany13,Ahmad13,El-Labany14,Zaghbeer14,Mamun15a,Guo16,Amina17a,Amina17b,Chowdhury17,Rahman18,Jahan19,EL-Shamy19,Shikha19,Sumi19a,Khaled19,Jahan20}  to investigate nonlinear structures (viz. solitary and rogue waves, shock structures, double layers, etc.) associated with dust-acoustic (DA) waves in such a novel OPDP medium during the last two
decades. These works are valid only when the effect of the self-gravitational force is negligible in comparison to the electrostatic force or the effect of self-gravitational force is insignificant. But the effect of self-gravitational force
increases as the size of OPDS increases, and begins to play the significant role on the dynamics of the OPDS. It is now well established that on the dynamics of OPDS in OPDP system with $r_d > 1\mu$m (where $r_d$ is the dust grain radius),
the effect of gravitational force is comparable to or greater than that of the electrostatic force \cite{Chow93,Medis94,Medis04,Shukla06,Howard99,Howard00}, and that the dust sizes vary from $1 \mu$m to $15 \mu$m in the most space and laboratory dusty plasma systems \cite{Havnes96,Tsintikidis96,Horanyi96,Gelinas98,Ali98,DAngelo01,DAngelo02,Chow93,Medis94,Medis04,Shukla06}. Thus, the dynamics of the OPDS in any OPDP system with $r_d > 1\mu$m  is governed  by the combined effects of electrostatic and self-gravitational forces. It means that the effect of the self-gravitational force cannot be neglected in any investigation on collective processes in such OPDP systems.

Mamun and Schlickeiser \cite{Mamun15b} first considered a self-gravitating OPDP system containing
cold OPDS and Boltzman distributed ion species (BDIS), which obeys the Boltzmann law, to study effects of self-gravitational field on the DA solitary waves \cite{Mamun15b}. Then \cite{Mamun15b} included the effect of strong co-relation among positive as well as negatively charged dust species, and studied the DA shock waves \cite{Mamun16}. The works of Mamun and Schlickeiser \cite{Mamun15b,Mamun16} have been further extended to different realistic situations of space and laboratory OPDP systems by a number authors, viz. Sabetkara and Dorranian \cite{Sabetkara16}, El-Taibany {\it et al.} \cite {El-Taibany19}, Sumi {\it et al.} \cite{Sumi19b}, etc. Sabetkara and Dorranian \cite{Sabetkara16} considered the self-gravitating OPDP system containing OPDS, kappa distributed ion species, and Boltzmann electron species to study the DA solitary waves and double layers. El-Taibany {\it et al.} \cite {El-Taibany19} assumed the self-gravitating OPDP system (containing inertial cold OPDS and q-distributed ion as well as electron species) in presence of external magnetic field and polarization force acting on OPDS to study the DA periodic and solitary waves. Sumi {\it et al.} \cite{Sumi19b} considered a self-gravitating dissipative OPDP system containing  trapped ion and Boltzmann electron species to investigate the effect of trapped ions on the DA shock waves.

The works \cite{Mamun15b,Mamun16,Sabetkara16,El-Taibany19,Sumi19b} discussed above have been performed by employing the reductive perturbation method with appropriate stretching of independent variables \cite{Mamun19}, which are valid for small amplitude DA solitary and shock waves in self-gravitating OPDP systems. To the best knowledge of the authors, no correct work,
which is valid for arbitrary value of DA electrostatic and self-gravitation potentials, are carried out so far. Therefore, in our present work, we consider the perturbed state of a self-gravitating OPDP system  containing  OPDS and Boltzmann ion species, derive two coupled second-order nonlinear differential equations (which are valid for arbitrary values of electrostatic and self-gravitational potentials associated with the DA waves), numerically solve them to estimate the values of these potentials, and to make a well-defined comparison between the electrostatic and self-gravitational wave potentials.

The manuscript is structured as follows. The basic equations governing the dynamics of nonlinear DA waves propagating in the perturbed state of a self-gravitating OPDP system are given in Sec. II. Two coupled second-order nonlinear differential equations (which are valid for arbitrary values of the electrostatic and self-gravitational potentials associated with the DA waves) are derived and numerically solved them to estimate and compare the values of these potentials for two opposite situations (viz. $\varrho<1$ and $\varrho>1$)  in Sec. III. A brief discussion is finally provided in Sec.  IV.

\section{Governing Equations}
We consider a self-gravitating OPDP system containing the OPDS and BDIS. Thus, at equilibrium,
we have $Z_{+}N_{+}^0 + N_{i}^0 = Z_{-}N_{-}^0$, where $N_{+}^0$ ($N_{-}^0$) and $N_{i}^0$ are the positive (negative) dust number density, and BDIS number density, respectively. The perturbed state of such a OPDP system is governed by
\begin{eqnarray}
&&\frac{\partial \tilde{N}_{\pm}}{\partial t}+ \frac{\partial}{\partial x}[(\tilde{N}_{\pm}+1)U_{\pm}] = 0\,,
\label{cont}\\
&&\frac{\partial U_{\pm}}{\partial t} + U_{\pm}\frac{\partial U_{\pm}}{\partial x} = -S_{\pm}\frac{\partial \Phi}{\partial x} - \frac{\partial \Psi}{\partial x}\,,
\label{momen}\\
&&\frac{\partial^2 \Phi}{\partial x^2} = \tilde{N}_{-} - \alpha \tilde{N}_{+}- (1-\alpha)({\cal W}_i-1)\,,
\label{poisson-e}\\
&&\frac{\partial^2 \Psi}{\partial x^2} = \gamma(\tilde{N}_{-} + \beta\tilde{N}_{+})\,,
\label{poisson-g}
\end{eqnarray}
where the upper (lower) subscript $+$ ($-$) stands for positive (negative) dust species; $\tilde{N}_{+}$ ($\tilde{N}_{-}$) is the perturbed part of the positive (negative) dust number density normalized by its equilibrium value $N_{+}^{0}$ ($N_{-}^{0}$); $U_{+}$ ($U_{-}$) is positive (negative) dust fluid speed normalized by $C_{d-}=(Z_{-}k_BT_i/m_{-})^{1/2}$ ($T_i$ being the ion temperature and $k_B$ being the Boltzmann constant); $\Phi$ is the DA wave electrostatic potential normalized by $k_BT_i/e$; $\Psi$ is the DA wave self-gravitational potential normalized by $C_{d-}^2$; ${\cal W}_i=\exp(-\Phi)$ represents the BDIS  number density normalized to its equilibrium $N_i^0$; $x$ is the space variable normalized by $\lambda_{Dm}=(k_BT_i/4\pi Z_{-}N_{-}^{0}e^2)^{1/2}$,  $t$ is the time variable normalized by the inverse of the negatively charged dust plasma
frequency $\omega_{p-}^{-1}=(m_{-}/4\pi Z_{-}^2N_{-}^{0}e^2)^{1/2}$; $S_+ = \mu\left(= Z_{+}m_{-}/Z_{-}m_{+}\right)$ and $S_{-}=-1$; $\alpha = Z_+N_{+}^0/Z_{-}N_{-}^0$, $\beta= N_{+}^0m_+/N_{-}^0m_-$, and $\gamma=\omega_{J-}^2/\omega_{p-}^2$ [with $\omega_{J-} = (4\pi GN_{-}^{0}m_{-})^{1/2}$ being the Jeans frequency for the negative dust species and $G$ being the universal gravitational constant]. The basic equations describing the perturbed state of the OPDP system under consideration can be interpreted as follows.
\begin{itemize}
\item{Equation (\ref{cont}) is the continuity equation for perturbed positive and negative dust fluids, where effects of source and sink terms have been neglected.}
\item{Equation (\ref{momen}) represents the momentum equations for positive and negative dust fluids, where the effects of electrostatic and self-gravitation forces are included. These are provided by the first and second terms on the right-hand side of it.}
\item{Equation (\ref{poisson-e}) represents Poisson's equation for the perturbed electrostatic potential. The last term of it represents the perturbed part of the number density of the BDIS.}
\item{Equation (\ref{poisson-g}) represents Poisson's equation for perturbed self electrostatic potential. It is important to note here that the effect of the self-gravitational force on ion species has been neglected since the mass number density of the ion species is neglected in comparison with that of the positive or negative dust species.}
\end{itemize}
We note that we have neglected the contribution of the free electrons since the latter were almost completely
depleted to the surface of the negative dust species, and that we have started with the perturbed state of the OPDP system,
since we interested in examining wave electrostatic and self-gravitational potential structures, and since Poisson's equation for the self-gravitational potential cannot be defined at equilibrium state of the OPDP system under present consideration.

\section{Electrostatic and Self-gravitational Potentials}
We first assume that all the dependent variables depend only on a single variable $\xi = x - M\tau$ with $\tau=t$, where $M$ is the DA wave speed normalized by $C_{d-}$, and express (\ref{cont})$-$(\ref{poisson-g}) as
\begin{eqnarray}
&&\frac{\partial \tilde{N}_{\pm}}{\partial \tau} - M \frac{\partial \tilde{N}_{\pm}}{\partial \xi} + \frac{\partial}{\partial \xi}\left[(\tilde{N}_{\pm}+1)U_{\pm}\right]=0\,,\label{1}\\
&&\frac{\partial U_{\pm}}{\partial \tau} - M \frac{\partial U_{\pm}}{\partial \xi} + U_{\pm}\frac{\partial U_{\pm}}{\partial \xi}= -S_{\pm}\frac{\partial \Phi}{\partial \xi}-\frac{\partial \Psi}{\partial \xi}\,,\label{2}\\
&&\frac{\partial^2 \Phi}{\partial \xi^2} =\tilde{N}_{-}-\alpha \tilde{N}_{+}-(1-\alpha)({\cal W}_i-1)\,,\label{3}\\
&&\frac{\partial^2 \Psi}{\partial \xi^2} = \gamma(\tilde{N}_{-}+\beta \tilde{N}_{+})\,.
\label{4}
\end{eqnarray}
Now, using (\ref{1}) and (\ref{2}) along with the steady state ($\partial/\partial\tau \rightarrow 0$) and boundary conditions ($\tilde{N}_{\pm} \rightarrow 0,\,\,U_{\pm} \rightarrow 0$, $\Phi \rightarrow 0$, and $\Psi \rightarrow 0$ at $\xi \rightarrow \pm \infty$), one can express $\tilde{N}_{\pm}$ as
\begin{equation}\label{N}
\tilde{N}_{\pm} = \frac{U_{\pm}}{M - U_{\pm}}\,,
\end{equation}
where
\begin{equation}\label{U}
U_{\pm} = M\left[1 - \sqrt{1 - \frac{2(S_{\pm}\Phi + \Psi)}{M^2}}\right]\,.
\end{equation}
Equation \eqref{U} yields
\begin{eqnarray}
 &&\tilde{N}_{+} ={\cal W}_{+}-1\,,
 \label{N-p}\\
 &&\tilde{N}_{-} ={\cal W}_{-}-1\,,
 \label{N-m}
\end{eqnarray}
where
\begin{eqnarray}
&&{\cal W}_{+}=\left[1- \frac{2}{M^2}(\mu\Phi+\Psi)\right]^{-\frac{1}{2}}\,,\\
&&{\cal W}_{-}=\left[1 + \frac{2}{M^2}(\Phi  - \Psi)\right]^{-\frac{1}{2}}\,.
\end{eqnarray}
Now, substituting (\ref{N-p}) and (\ref{N-m})  into  (\ref{3}) and (\ref{4}), we obtain two coupled equations in the form
\begin{eqnarray}
 &&\hspace*{-10mm}\frac{d^2 \Phi}{d \xi^2} =({\cal W}_{-}-1)-\alpha({\cal W}_{+}-1)- (1-\alpha)({\cal W}_i-1),
 \label{coupled-el}\\
 &&\hspace*{-10mm}\frac{d^2 \Psi}{d \xi^2} =\gamma[({\cal W}_{-}-1) + \beta({\cal W}_{+}-1)]\,.
 \label{coupled-em}
 \end{eqnarray}
We now numerically analyze the two coupled equations represented by (\ref{coupled-el}) and (\ref{coupled-em}) to estimate the electrostatic and self-gravitational potentials associated with the DA waves in the self-gravitating OPDP system in two opposite situations, namely $\varrho<1$ and  $\varrho>1$.
\subsection{${\bf \varrho<1}$}
We first consider the OPDP system, where $\varrho<1$. The latter indicates that the size (which significantly increases the mass and the magnitude of dust charge) of the positively charged dust species is smaller than that of the negatively charged species.
We have first numerically analyzed (\ref{coupled-el}) and (\ref{coupled-em}) to estimate the electrostatic and self-gravitational potentials ($\Phi$ and $\Psi$) and to compare the value of one with that of the other in the self-gravitating OPDP system in the situation of $\varrho<1$.  The numerical results are displayed in figures \ref{Fig1}-\ref{Fig3}, where we have used exactly the same set of OPDP parameters (viz. $\alpha=0.1$-$0.6$, $\beta=0.1$, $\mu=0.1$, $M=1.1$ and $\gamma=0.7$) which satisfy $\varrho<1$. We note that $\gamma$ and $M$ is independent of the size of the positive dust species, and have no any role in violating the condition $\varrho<1$.

It is important to mention that in the left panels of figures \ref{Fig1} and \ref{Fig2}, we have numerically integrated (\ref{coupled-el}) and (\ref{coupled-em}) from $\xi =0$ to $\xi=-\infty$ with the conditions $\Phi =\Phi_m$ and $\Psi =\Psi_m$ [where $\Phi_m$ ($\Psi_m$) is the maximum value of the electrostatic (self-gravitational) potentials in its profile] at $\xi=0$, and that in the right panels of figures \ref{Fig1} and \ref{Fig2}, we have numerically integrated (\ref{coupled-el}) and (\ref{coupled-em}) from $\xi =0$ to $\xi=\infty$ with the same conditions (viz.  $\Phi =\Phi_m$ and $\Psi =\Psi_m$ at $\xi=0$). It is obvious from figures \ref{Fig1} and \ref{Fig2} that the variation of both electrostatic and self-gravitational) potentials with space variable ($\xi$) represent the DA solitary like structures associated with $\Phi>0$ and $\Psi<0$. They imply that both amplitude and width of these DA solitary like structures increase with the rise of the value of $\alpha$. On the other hand, figure \ref{Fig3} represents a well-defined comparison between the electrostatic and self-gravitational wave potentials ($\Phi$ and $\Psi$). It indicates that  the self-gravitational potential ($\Psi$) increases with the increase in the electrostatic potential ($\Phi$), and that as the value of $\alpha$ increases the self-gravitational potential ($\Psi$) and the electrostatic potential ($\Phi$) decrease. It is also seen that the electrostatic wave potential ($\Phi$)  is approximately twenty times larger than that of the self-gravitational wave potential ($\Psi$). The solid (corresponding to $\alpha=0.1$), dotted ($\alpha=0.3$), dashed ($\alpha=0.5$) and dot-dashed ($\alpha = 0.6$) curves end at $\Psi\simeq -0.0175$, $\Psi\simeq -0.014$, $\Psi\simeq -0.01$, and $\Psi\simeq -0.0085$ respectively. This is due to fact that the upper limit for the value of the self-gravitational potential reduces as the value of $\alpha$ increases, and after the end points of these curves, the electrostatic force completely dominated over the gravitational force. The OPDP parameters used in the numerical analysis depicted in figures \ref{Fig1}$-$\ref{Fig3} correspond to any OPDP system, where $\varrho<1$, i.e. the size of the positively charged dust species is smaller than that of the negatively charged dust species.
\subsection{${\bf \varrho>1}$}
We now consider the situation opposite to that considered in {\bf subsection A}, i.e. $\varrho>1$ indicating that the size of the positively charged dust species is larger than that of the negatively charged dust species. We have again numerically analyzed (\ref{coupled-el}) and (\ref{coupled-em}) to estimate the electrostatic and self-gravitational potentials ($\Phi$ and  $\Psi$) and to compare the value of one with that of the other
\begin{figure*}[htb!]
\centering
\begin{tabular}{@{}cc@{}}
   \includegraphics[width=0.48\textwidth]{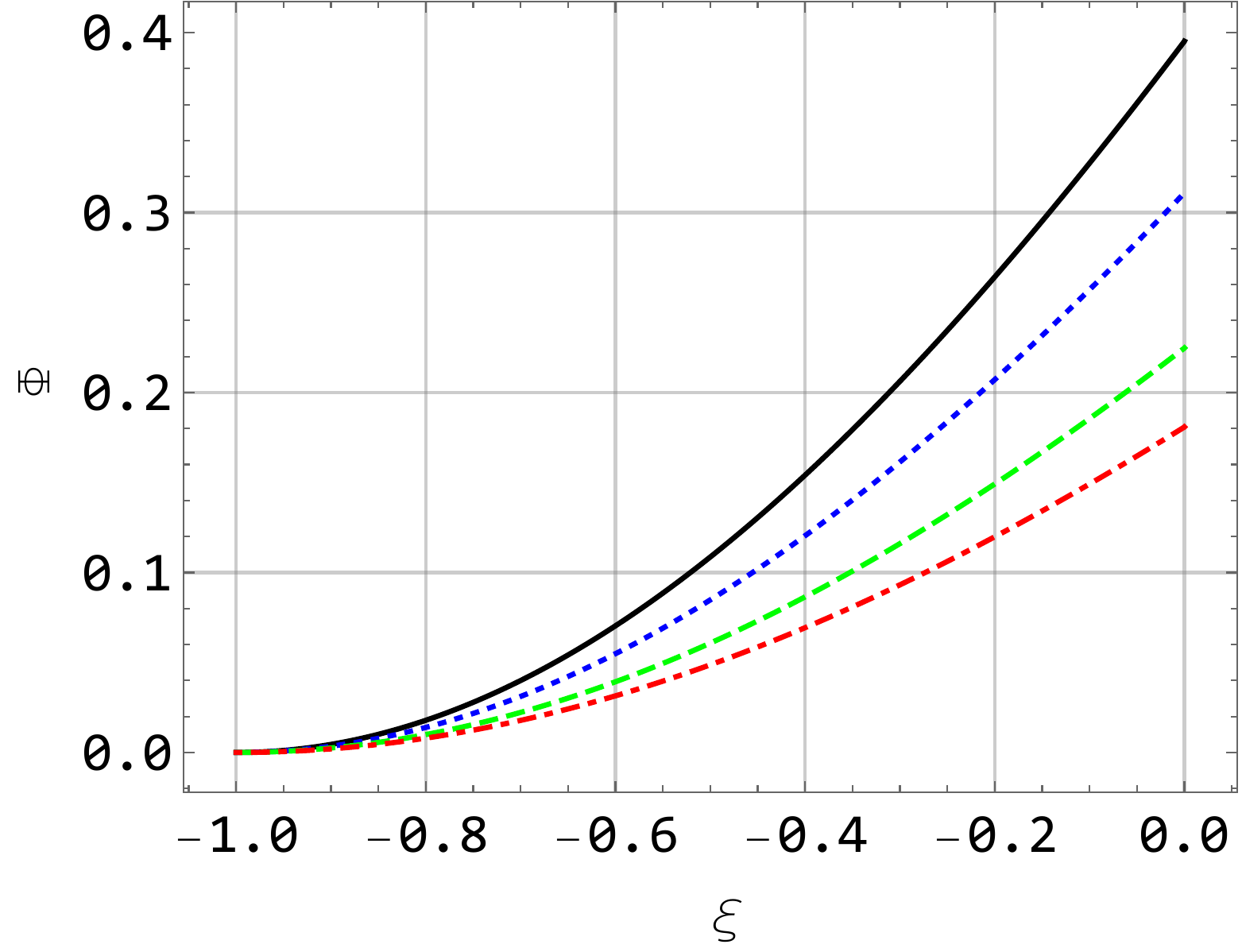} &
   \includegraphics[width=0.48\textwidth]{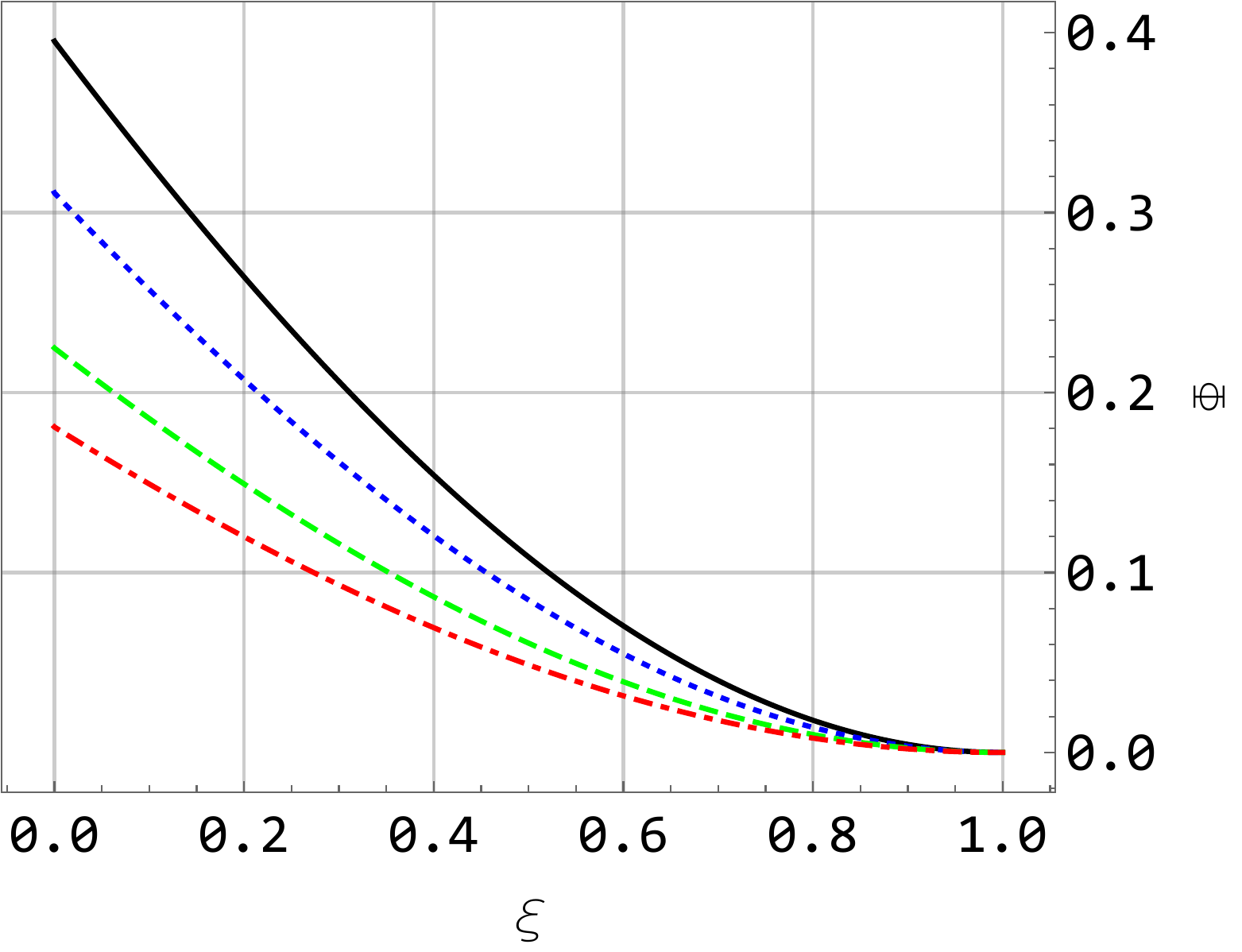}
\end{tabular}
\caption{The variation of $\Phi$ with $\xi$ in $-\xi$-axis (left panel) and in $+\xi$-axis (right panel) for $\beta=0.1$, $\mu=0.1$, $\gamma=0.7$, $M=1.1$, $\alpha=0.1$ (solid curve), $\alpha=0.3$ (dotted curve), $\alpha =0.5$ (dashed curve), and $\alpha=0.6$ (dot-dashed curve).}
\label{Fig1}
\end{figure*}
\begin{figure*}[htb!]
\centering
\begin{tabular}{@{}cc@{}}
   \includegraphics[width=0.48\textwidth]{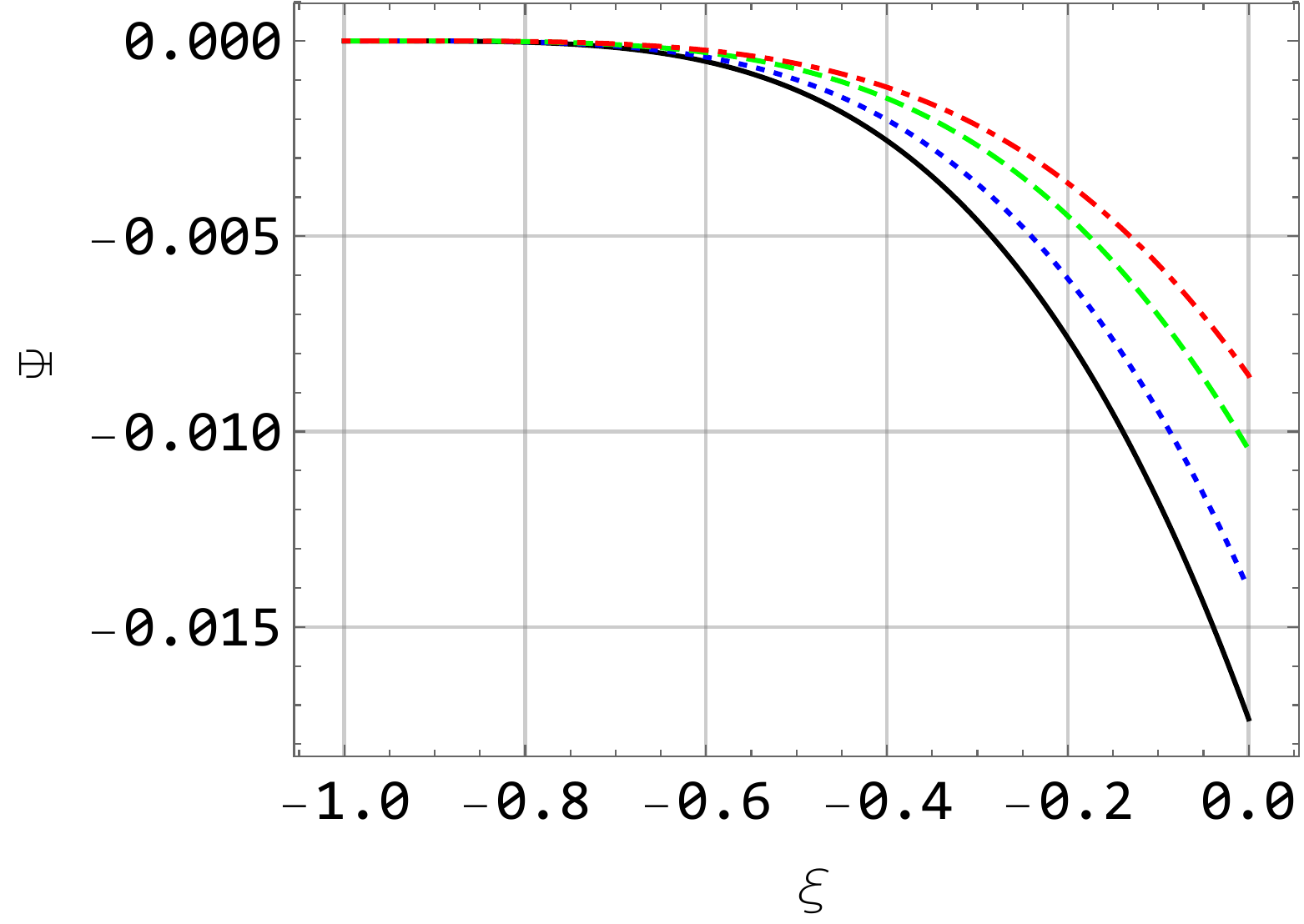} &
   \includegraphics[width=0.48\textwidth]{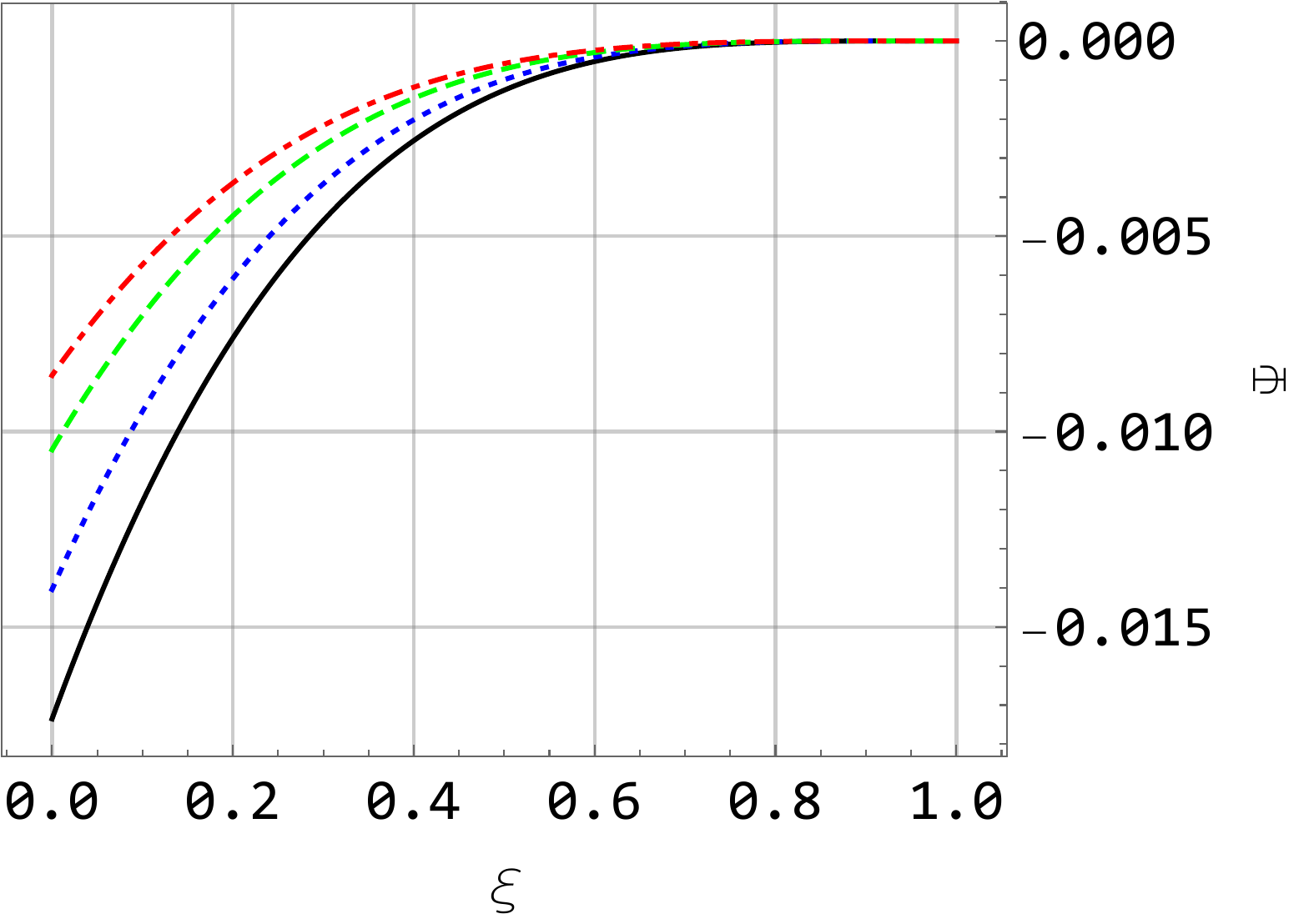}
\end{tabular}
\caption{The variation of $\Psi$ with $\xi$ in $-\xi$ axis (left panel) and in $+\xi$ axis (right panel) for $\beta=0.1$, $\mu=0.1$, $\gamma=0.7$, $M=1.1$, $\alpha=0.1$ (solid curve), $\alpha=0.3$ (dotted curve), $\alpha =0.5$ (dashed curve), and $\alpha=0.6$ (dot-dashed curve).}
\label{Fig2}
\end{figure*}
\begin{figure}[H]
\centering
\includegraphics[width=80mm]{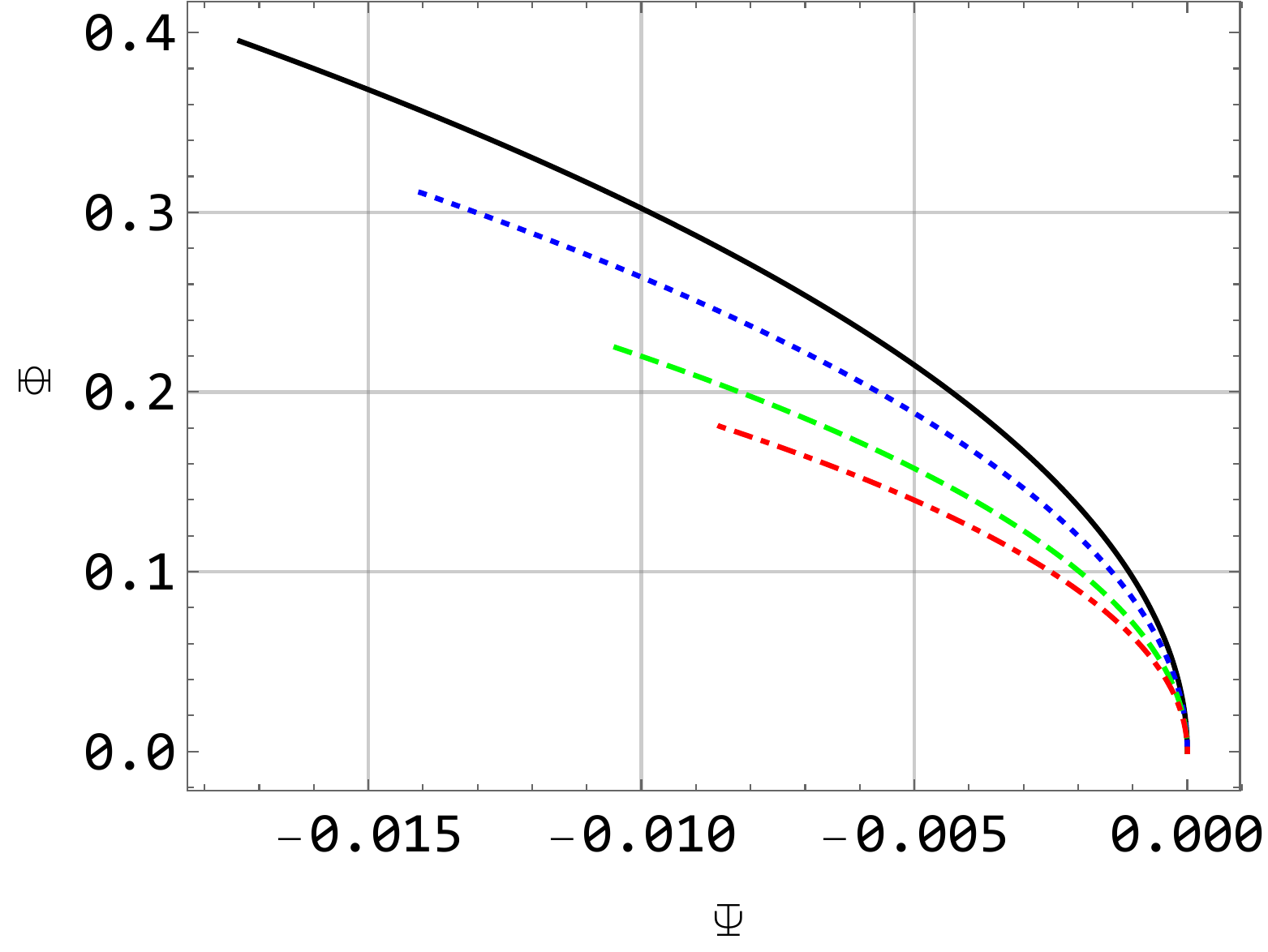}
\caption{The variation of $\Phi$ with $\Psi$ for $\beta=0.1$, $\mu=0.1$, $\gamma=0.7$, $M=1.1$, $\alpha=0.1$ (solid curve), $\alpha=0.3$ (dotted curve), $\alpha =0.5$ (dashed curve), and $\alpha=0.6$ (dot-dashed curve).}
\label{Fig3}
\end{figure}
\noindent in the situation of $\varrho>1$. The numerical results are depicted in figures \ref{Fig4}--\ref{Fig6}, where we have used exactly the same set of OPDP parameters (viz. $\alpha=0.1$-$0.6$, $\beta=2.0$, $\mu=1.1$, $M=1.1$, and $\gamma=0.7$) which satisfy $\varrho>1$. We again note that $\gamma$ and $M$ is independent of the size of the positive dust species, and have no any role in violating the condition $\varrho>1$. We also  note that our OPDP system under consideration does not allow $Z_{+}N_{+}^{0} > Z_{-}N_{-}^{0}$ (i.e. $\alpha > 1$) which breaks down the quasi-neutrality. Therefore, the DA wave electrostatic potential with $\Phi < 0$ has not been observed. It is important to mention that in the left panels of figures \ref{Fig4} and \ref{Fig5}, we have numerically integrated (\ref{coupled-el}) and (\ref{coupled-em}) from $\xi =0$ to $\xi=-\infty$ with the conditions $\Phi =\Phi_m$ and $\Psi =\Psi_m$ at $\xi=0$, and that in the right panels of figures \ref{Fig4} and \ref{Fig5}, we have numerically integrated (\ref{coupled-el}) and (\ref{coupled-em}) from $\xi =0$ to $\xi=\infty$ with the same conditions (viz. $\Phi =\Phi_m$ and $\Psi =\Psi_m$ at $\xi=0$).
\begin{figure*}[htb!]
\centering
\begin{tabular}{@{}cc@{}}
   \includegraphics[width=0.48\textwidth]{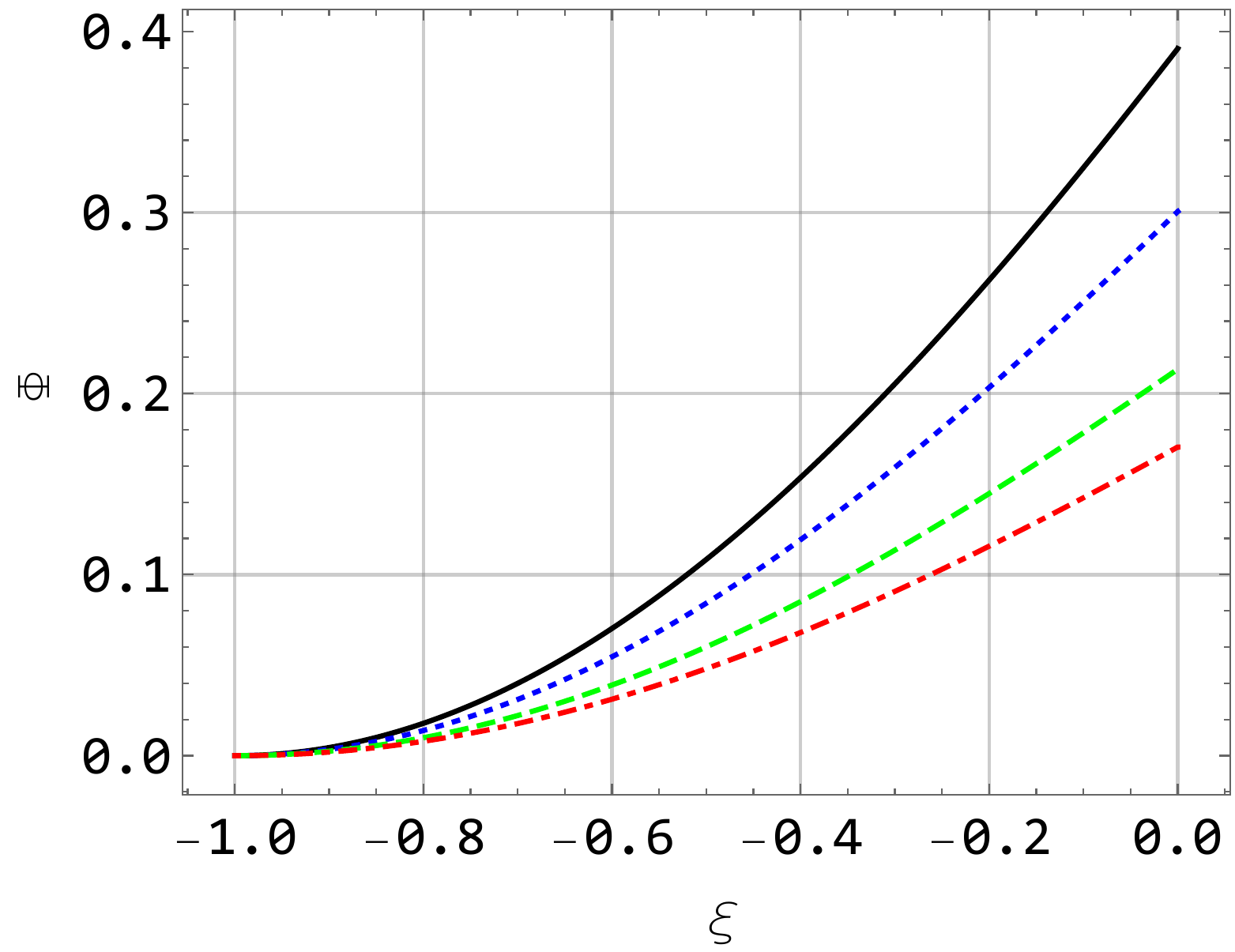} &
   \includegraphics[width=0.48\textwidth]{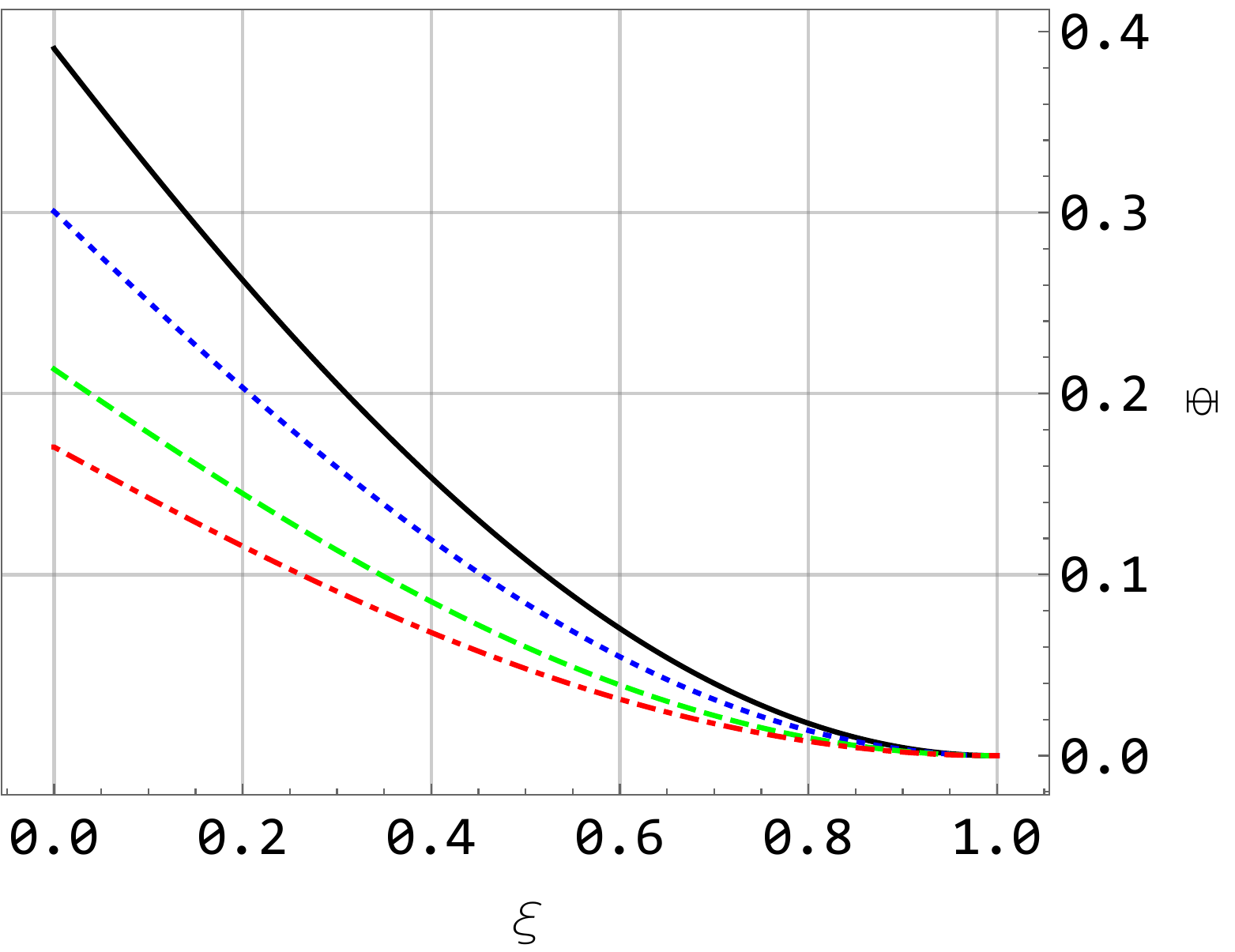}
\end{tabular}
\caption{The variation of $\Phi$ with $\xi$ in $-\xi$ axis (left panel) and in $+\xi$ axis (right panel) for $\beta=2.0$, $\mu=1.3$, $\gamma=0.7$, $M=1.1$, $\alpha=0.1$ (solid curve), $\alpha=0.3$ (dotted curve), $\alpha =0.5$ (dashed curve), and $\alpha=0.6$ (dot-dashed curve).}
\label{Fig4}
\end{figure*}
\begin{figure*}[htb!]
\centering
\begin{tabular}{@{}cc@{}}
   \includegraphics[width=0.48\textwidth]{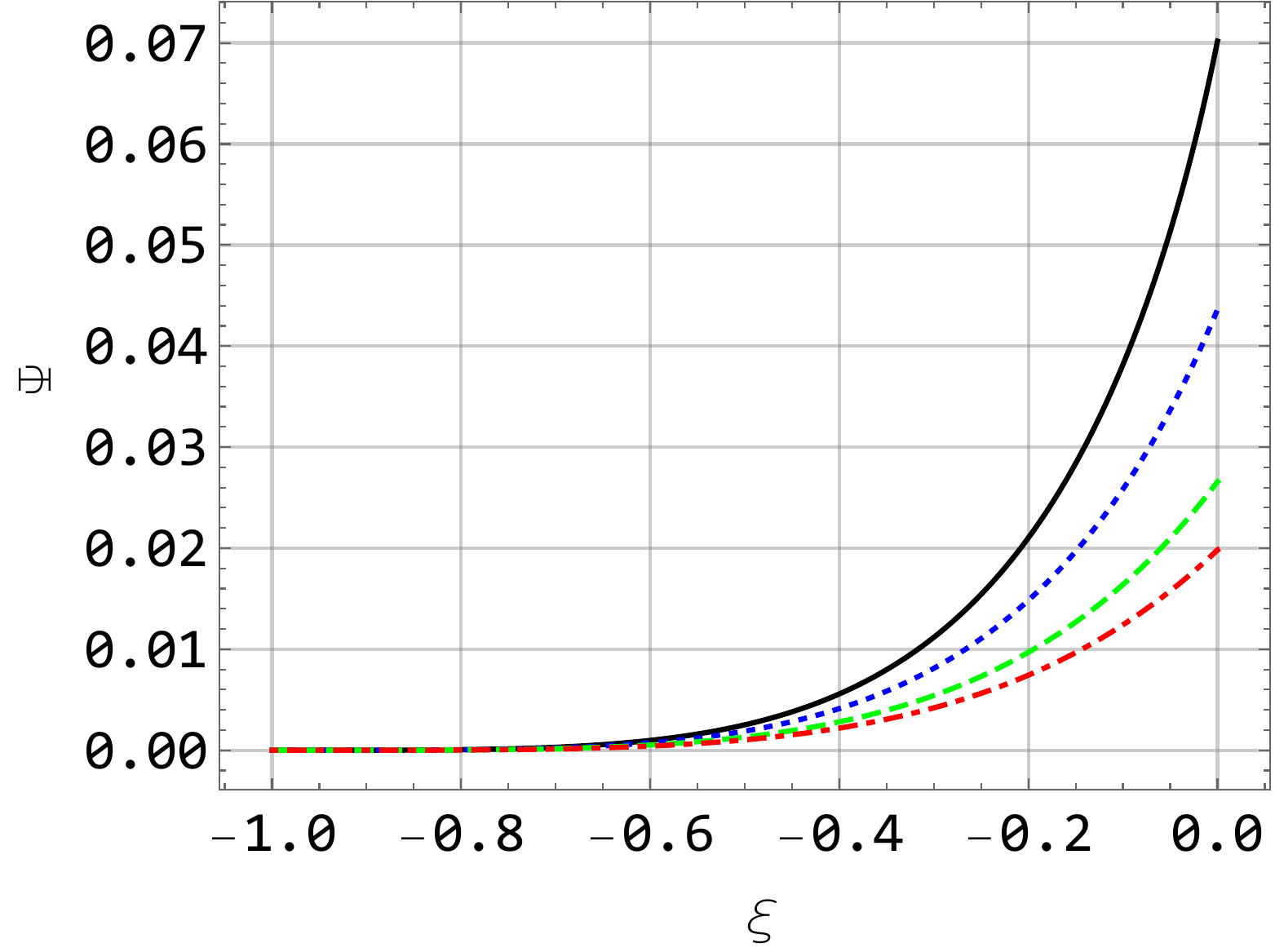} &
   \includegraphics[width=0.48\textwidth]{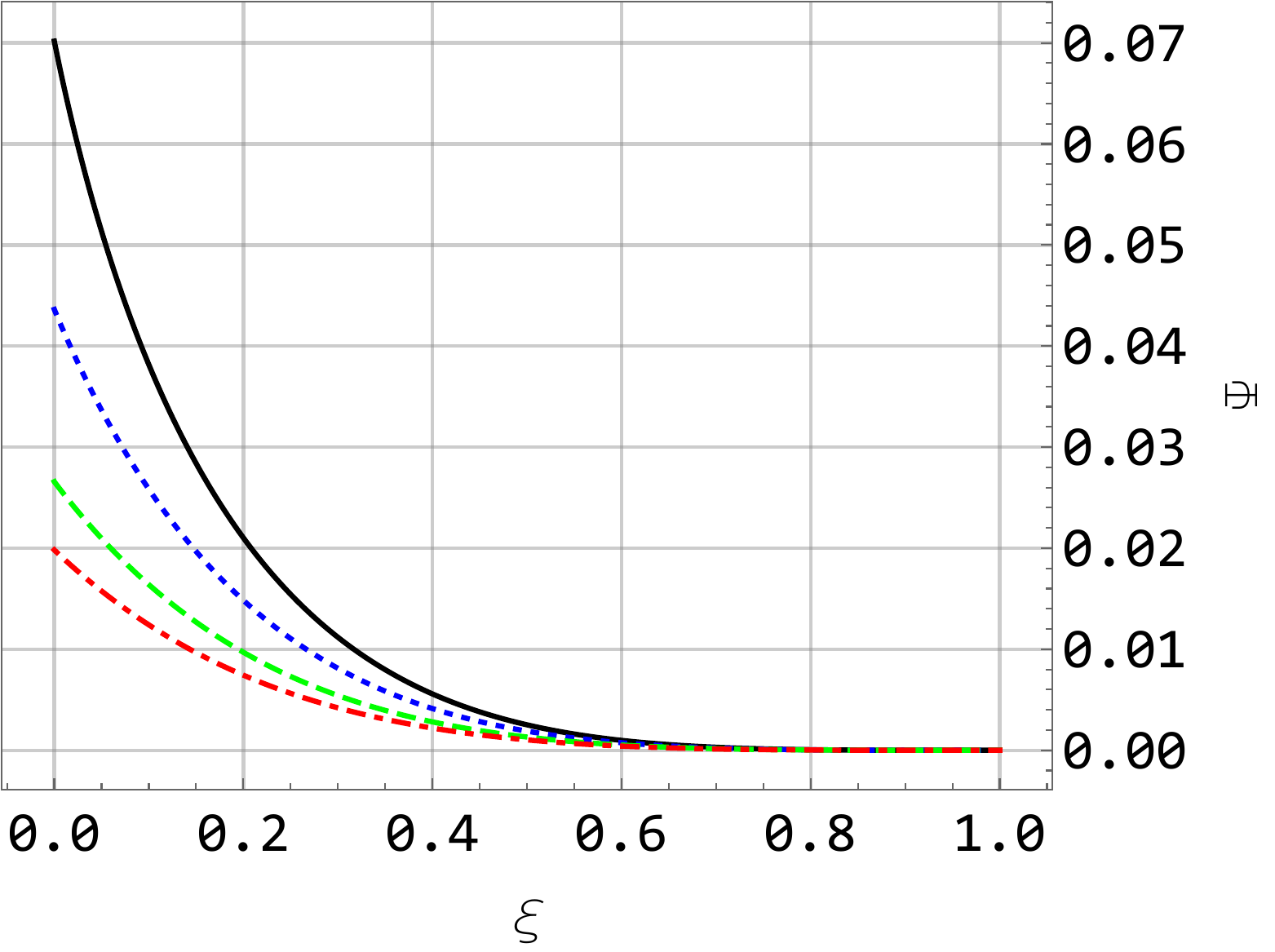}
\end{tabular}
\caption{The variation of $\Psi$ with $\xi$ in $-\xi$ axis (left panel) and in $+\xi$ axis (right panel) for $\beta=2.0$, $\mu=1.3$, $\gamma=0.7$, $M=1.1$, $\alpha=0.1$ (solid curve), $\alpha=0.3$ (dotted curve), $\alpha =0.5$ (dashed curve), and $\alpha=0.6$ (dot-dashed curve).}
\label{Fig5}
\end{figure*}
\begin{figure}[htb!]
\centering
\includegraphics[width=82mm]{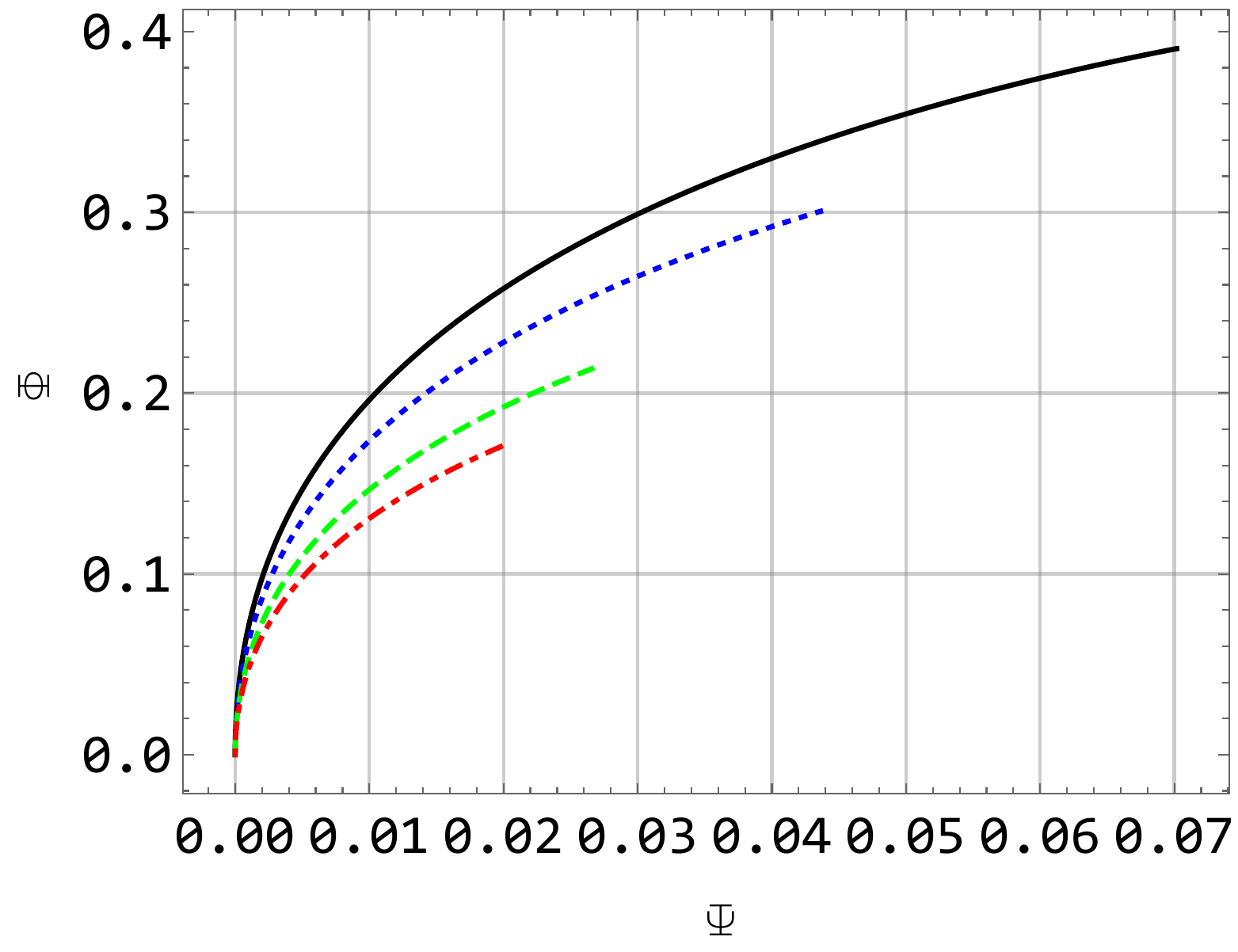}
\caption{The variation of $\Phi$ with $\Psi$ for $\beta=2.0$, $\mu=1.3$, $\gamma=0.7$, $M=1.1$, $\alpha=0.1$ (solid curve), $\alpha=0.3$ (dotted curve), $\alpha =0.5$ (dashed curve), and $\alpha=0.6$ (dot-dashed curve).}
\label{Fig6}
\end{figure}
It is obvious from figure \ref{Fig4} (\ref{Fig5}) that the variation of the electrostatic (self-gravitational) potential with space
variable ($\xi$) represents the DA solitary like structures associated with $\Phi>0$ ($\Psi>0$). They imply that
the amplitude and width of both electrostatic and self-gravitational solitary like structures decrease with the increase in the value of $\alpha$. On the other hand, \ref{Fig6} represents a well-defined comparison between the electrostatic and self-gravitational wave potentials ($\Phi>0$ and $\Psi>0$). It indicates that the self-gravitational potential ($\Psi$) increases with the increase in the electrostatic potential ($\Phi$), and that as the value of $\alpha$ increases the self-gravitational potential ($\Psi$) and the electrostatic potential($\Psi$) decreases.  It is also seen that the magnitude of the electrostatic wave potential ($\Phi$) is approximately five times larger than that of the self-gravitational wave potential ($\Psi$). The solid (corresponding to $\alpha=0.1$), dotted (corresponding to $\alpha=0.3$), dashed ($\alpha=0.5$), and dot-dashed ($\alpha=0.6$) curves end at $\Psi\simeq 0.07$, $\Psi\simeq 0.044$, $\Psi\simeq 0.027$, and $\Psi\simeq 0.02$, respectively. This is due to fact that the upper limit for the value of the self-gravitational potential reduces as the value of $\alpha$ increases, and after the end points of these curves, the electrostatic force completely dominated over the gravitational force. The OPDP parameters used in the numerical analysis depicted in figures \ref{Fig4}$-$\ref{Fig6} correspond to any OPDP system, where $\varrho>1$, i. e. the size of the positively charged dust species is larger than that of the negatively charged dust species. It is found here that the present situation ($\varrho>1$) of OPDP system supports the DA self-gravitational solitary like structures with $\Psi>0$, but the opposite situation ($\varrho<1$) considered in {\bf subsection A} supports the DA self-gravitational solitary like structures with $\Psi<0$. It is important to note that any OPDP system satisfying either $\varrho<1$ or $\varrho>1$ supports the DA wave electrostatic solitary like structures with $\Phi>0$. It is observed from figures \ref{Fig1}$-$\ref{Fig6} that DA wave electrostatic potential decreases with $\mu$, but it does not change with $\beta$ and $\gamma$. On the other hand, the DA self-gravitational potential increases rapidly with $\beta$ and $\mu$. The increase in $\gamma$ causes to increase (decrease) the amplitude of DA self-gravitational solitary like structures with $\Psi > 0$ ($\Psi <0$). As we increase $\beta$ and $\mu$, the self-gravitational forces completely dominate over the electrostatic forces since $\alpha$ is always less than $1$.

\section{Discussion}
The self-gravitating OPDP system  containing  OPDS and BDIS has been considered, and the basic features (viz. polarity, amplitude, and width) of the solitary like structures associated with the DA wave electrostatic and self-gravitational potentials
have been examined by the numerical analysis of two coupled  second-order nonlinear differential equations. The latter have been derived  from  the continuity and momentum equations for the  positive and  negative dust species, and the Boltzmann law for the ion species.  The results obtained from the numerical analysis,  which has been been displayed by the figures
\ref{Fig1}$-$\ref{Fig6}, can be pinpointed as follows:
\begin{itemize}
\item{The DA wave electrostatic and self-gravitational potentials ($\Phi$ and $\Psi$) decrease with the increase in the value of space variable $\xi$ in the form of solitary-like structure, and that $\Phi\rightarrow 0$ and $\Psi\rightarrow 0$ at $\xi\rightarrow \pm\infty$,  and $|\Phi|$ and $\Psi$ are maximum at $\xi=0$.}

\item{The OPDP system with $\varrho<1$, i.e. with the size of positively charged dust species is smaller than that of negatively charged dust species, supports the DA solitary structures with $\Phi>0$ and $\Psi<0$, but the system with the opposite situation ($\varrho<1$, i.e. the size of positively charged dust species is larger than that of negatively charged dust species) the DA solitary structures with $\Phi>0$ and $\Psi>0$. This means that any OPDP system satisfying either $\varrho<1$ or $\varrho>1$ supports the electrostatic solitary structures with $\Phi>0$.}

\item{The magnitude of the DA electrostatic and self-gravitational potentials ($\Phi$ and $\Psi$) increase with the decrease in
$\alpha$.}

\item{The DA wave electrostatic potential is much larger than the DA wave self-gravitational potential. However, the DA wave self-gravitational potential becomes larger than the DA wave electrostatic potential with the rise of the values of $\beta$. This is due to the facts that the masses of the OPDS are assumed to be so massive that the self-gravitational force dominates over the electrostatic force.}

\item{The wave self-gravitational potential increases with the increase in the wave electrostatic potential. This physics of it is that both the mass and the magnitude of charge of both dust species increase as the size of the dust species increases, and consequently, the magnitude of charge of the dust species increases as well as their mass increases.}
\end{itemize}
To conclude, we have first time correctly identified the basic features of arbitrary values of electrostatic and gravitational potentials associated with the DA waves in OPDP systems by numerical analysis of two coupled nonlinear differential equations derived from the continuity and momentum equations for positive and negative dust species, and Boltzmann ion species. We, therefore, hope that our present investigation is useful in understand the physics of localized electrostatic and self-gravitational disturbances in space environments (viz. Earth's mesosphere \cite{Havnes96,Gelinas98}, cometary tails \cite{Horanyi96}, Saturn ring \cite{Tsintikidis96,Horanyi96}, etc.) and different laboratory devices  \cite{Ali98,DAngelo01,DAngelo02}.
\vspace{0.5cm}
\acknowledgements
A. Mannan gratefully acknowledges the financial support of the Alexander von Humboldt Stiftung (Bonn, Germany) through its post-doctoral research fellowship.

\end{document}